\documentclass[preprint,showpacs,preprintnumbers,amsmath,amssymb]{revtex4}

\usepackage[dvips]{graphicx}
\usepackage{epsfig}
\usepackage{amsmath}
\usepackage{graphics}

\newcommand{\coo}{{\bf r}, t}

\begin{document}

\title{Dynamic Structure Factor of Normal Fermi Gas 
\\ from Collisionless to Hydrodynamic Regime} 

\author{Shohei Watabe$^{1}$} 
\altaffiliation{Present Address: 
Department of Physics, Keio University, 3-14-1 Hiyoshi, Kohoku-ku, Yokohama 223-8522, Japan}
\altaffiliation{CREST(JST), 4-1-8 Honcho, Kawaguchi, Saitama 332-0012, Japan}
\author{Tetsuro Nikuni$^{2}$}
\affiliation{
$^{1}$
Institute of Physics, Department of Physics, The University of Tokyo, Komaba 3-8-1 Meguro-ku, Tokyo 153-8902, Japan
}
\affiliation{
$^{2}$
Department of Physics, Tokyo University of Science, 
1-3 Kagurazaka, Shinjuku-ku, Tokyo 162-8601, Japan
}

\begin{abstract} 
The dynamic structure factor of a normal Fermi gas is investigated by using the moment method 
for the Boltzmann equation.
We determine the spectral function at finite temperatures 
over the full range of crossover from the collisionless regime to the hydrodynamic regime. 
We find that the Brillouin peak in the dynamic structure factor 
exhibits a smooth crossover from zero to first sound as functions of temperature and interaction strength. 
The dynamic structure factor obtained using the moment method 
also exhibits a definite Rayleigh peak ($\omega \sim 0$), 
which is a characteristic of the hydrodynamic regime. 
We compare the dynamic structure factor obtained by the moment method 
with that obtained from the hydrodynamic equations. 
\end{abstract}


\maketitle 
\section{Introduction}
When discussing collective modes in quantum many-body systems at finite temperatures, 
there are two regimes of interest: 
the collisionless (or mean-field) regime and the collisional (or hydrodynamic) regime. 
The mechanisms for the occurrence of collective modes in these two regimes differ critically. 
One of the collective modes is the first sound in the collisional regime, 
which is due to local equilibrium. 
In a normal Fermi system, 
zero sound is a characteristic sound mode in the collisionless regime.
It propagates due to the mean-field interaction. 
This zero sound was first predicted by Landau~\cite{Landau1957} based on Fermi liquid theory~\cite{Landau19561957}, 
and it has been studied in many fields of physics including low-temperature physics, nuclear physics (hot nuclear matter), and astrophysics (neutron stars).

Collective modes with time-dependent density disturbances have been investigated. 
The crossover between first sound and zero sound was first observed 
by Keen {\it et al.} in measurements of the acoustic impedance between liquid $^{3}$He and a quartz crystal~\cite{Keen1965}.  
Density fluctuations have been excited in ultracold atoms
by deforming the trapping potential~\cite{Andrews1997, Joseph2007}. 
An alternative way to probe collective excitations is to use scattering of lights or particles. 
This involves measuring the spectral function of the system, 
such as the dynamic structure factor or the density response function.

In neutron scattering experiments, 
the roton spectrum of liquid $^{4}$He was observed through the spectral function. 
In early experiments on liquid $^{3}$He, 
several problems were encountered in neutron scattering measurements of the dynamic structure factor 
including a high neutron absorption cross-section and interaction between spins. 
Nowadays, these problems have been overcome 
and the neutron scattering experiments on liquid $^{3}$He have also been performed~\cite{Scherm1974,Skold1976}. 
The dynamic structure factor of ultracold atoms has been studied by two-photon Bragg spectroscopy in a condensed Bose--Einstein gas and the Bogoliubov spectrum was obtained~\cite{Steinhauer2003}. 
This technique has also been used recently for a Fermi gas~\cite{Veeravalli2008}.

In 1958, Abrikosov and Khalatnikov conducted the first theoretical study of the dynamic structure factor of a normal Fermi liquid~\cite{Abrikosov1958}. They
proposed using light scattering
to observe the zero sound in liquid $^3$He. 
The spectral function has also been studied in connection with 
evaluating the Landau parameter by sum rules~\cite{Safier1972}. 
Photoabsorption cross-sections of hot nuclear matter
have been studied by 
taking only two moments: the density and current~\cite{Larionov1999,Baran1999}. 
In the field of the ultracold quantum gases, 
the spectral function of normal Fermi gases 
has been studied~\cite{Bruun2001,Stringari2009}. 

The dynamic structure factor of normal Fermi gases in both the collisionless and hydrodynamic regimes has been discussed in detail~\cite{Pines1994}. 
By employing the random-phase approximation, 
the dynamic structure factor can be discussed even beyond the phonon regime in the collisionless regime.
However, the hydrodynamic regime cannot be investigated using the same theoretical framework. 
In this regime, the hydrodynamic equations, which are not valid in the collisionless regime,
can be used to calculate the density response function.

Although the crossover between zero and first sound modes 
has been extensively studied for a long time, 
there has been no comprehensive study of the dynamic structure factor 
over the full crossover range from the collisionless regime to the collisional regime within a single theoretical framework. 
Furthermore, the dynamic structure factor for the crossover from zero to first sound has not been explicitly calculated 
and it is not obvious how it varies as a function of temperature and interaction strength. 
It is thus important to study the dynamic structure factor at finite temperatures from the collisionless regime to the hydrodynamic regime
within a single theoretical framework.

In the present paper, we study the dynamic structure factor of a normal Fermi gas
over the full crossover range from the collisionless regime to the hydrodynamic regime
within a single theoretical framework, namely the moment method. 
The moment method can be used to perform systematic analysis and it yields important physical insights. 
Gu\'ery-Odelin {\it et al.} first applied the moment method to study the collective mode in a trapped classical gas~\cite{Odelin1999A}. 
We recently used the moment method to study excitation spectra of a normal Fermi gas in a uniform system~\cite{Watabe2009}; 
the results for both frequency and damping of the collective mode clearly show the crossover from the zero to first sound mode. 
In the present paper, we extend the study of Ref.~\cite{Watabe2009} by investigating the dynamic structure factor over the full crossover range from zero to first sound.

This paper is organized as follows. 
Section~\ref{II} presents the moment method for a normal two-component Fermi gas. 
Section~\ref{III} examines the dynamic structure factor obtained using the moment method
and discusses the crossover from the collisionless to the hydrodynamic regime.
Section~\ref{IV} compares the spectral function obtained by the moment method 
with that obtained using the hydrodynamic equations.
We focus on the Brillouin peak in the dynamic structure factor, which is associated with the sound mode,
and the Rayleigh peak, which is associated with the thermal diffusion mode. 
We discuss the results in Sec.~\ref{V}. Section~\ref{VI} presents the conclusions. 

\section{Moment Equation and Dynamic Structure Factor}\label{II}
We start with the following Boltzmann equation: 
\begin{align}
&
\frac{\partial f_{\sigma}({\bf p}, \coo)}{\partial t}
+ 
\frac{\bf p}{m}\cdot \nabla_{\bf r} f_{\sigma}({\bf p}, \coo)
\nonumber
\\
&
\qquad -
\nabla_{\bf r}U_{\sigma}(\coo)
\cdot
\nabla_{\bf p}f_{\sigma}({\bf p}, \coo)
= 
{\mathcal I}_{\rm coll}
[f_{\sigma}], 
\label{Boltzmann}
\end{align}
where the subscript $\sigma =\{ \uparrow, \downarrow\}$ represents the spin component. 
We consider a normal Fermi gas with two spin components in 
the symmetric configuration $N_{\uparrow} = N_{\downarrow}$. 
We also assume that atoms with different spins collide with an $s$-wave scattering length of $a$. 
The effective potential $U_{\sigma}(\coo)$ is the sum of the mean-field interaction $g n_{-\sigma}(\coo)$ 
and the external field $U_{{\rm ext}}({\bf r}, t)$, where $n_{\sigma}(\coo)$ is the local density 
and $g$ is the interaction strength, which is given by $g = 4\pi\hbar^{2}a/m$. 
In this study, we consider a spin-independent external field.

We linearize the distribution function 
around the static equilibrium (denoted by $f_{\sigma}^{0} ({\bf p}, {\bf r})$)
using $f_{\sigma}({\bf p}, {\bf r}, t) 
= 
{f}_{\sigma}^{0} ({\bf p}, {\bf r}) + 
\delta f_{\sigma} ({\bf p}, {\bf r}, t)$. 
It is convenient to write fluctuations in the distribution function around 
the static equilibrium in terms of the average additional energy of the particles $\nu_{\sigma}({\bf p}, {\bf r}, t)$, 
which is defined by 
$ \delta f_{\sigma}({\bf p}, {\bf r}, t) 
\equiv 
(\partial f_{\sigma}^{0} / \partial \varepsilon_{\sigma}^{0} )
\nu_{\sigma}({\bf p}, {\bf r}, t)$. 
We apply a relaxation time approximation to the collision integral ${\mathcal I}_{\rm coll}[f_{\sigma}] $. 
Using this approximation, the collision integral can be reduced to 
\begin{eqnarray}	
{\mathcal I}_{\rm coll}[f_{\sigma}] 
&=&
-\frac{f_{\sigma}-\tilde{f}_{\sigma}}{\tau}
= 
-\frac{1}{\tau}
\frac{\partial f_{\sigma}^{0}}{\partial \varepsilon_{\sigma}^{0}}
\delta \nu_{\sigma}, 
\label{CollisionIntegral} 
\end{eqnarray}
where $\tau$ is the relaxation time, $\tilde{f}_{\sigma}$ is the distribution function in local equilibrium, 
and $\delta \nu_{\sigma} $ is given by 
$\delta \nu_{\sigma} = \nu_{\sigma} - \left [ A_{\sigma} +
{\bf B}  \cdot{\bf p} + C p^{2} \right ]$. 
$\nu_{\sigma, {\rm local}} \equiv A_{\sigma} + {\bf B}  \cdot{\bf p} + C p^{2}$ 
is the solution for the local equilibrium. 
The coefficients $A_{\sigma}$, ${\bf B}$, and $C$ are determined below using the conservation law.
We use the viscous relaxation time given in Ref.~\cite{Watabe2009} as the relaxation time $\tau$ in Eq.~(\ref{CollisionIntegral}).

We now consider an external field with the form 
$U_{\rm ext} ({\bf r}, t) = U_{\rm ext}({\bf q}, \omega) e^{i({\bf q}\cdot{\bf r} -\omega t)}$. 
This leads to a plane-wave solution of the linearized Boltzmann equation, which is
represented as 
$\nu_{\sigma} ({\bf p}, {\bf r}, t) = 
\nu_{\sigma}({\bf p}, {\bf q}, \omega) e^{i({\bf q}\cdot{\bf r} -\omega t)}$, and 
$\delta n_{\sigma} ({\bf r}, t) = \delta n_{\sigma}({\bf q}, \omega) e^{i({\bf q}\cdot{\bf r} -\omega t)}$. 
The linearized Boltzmann equation 
for $\nu_{\sigma}({\bf p})$ is now given by 
\begin{align}
&
\frac{\partial f_{\sigma}^{0}}{\partial \varepsilon_{\sigma}^{0}}
\left \{
\left (
\omega -\frac{{\bf p}\cdot{\bf q}}{m}
\right )
\nu_{\sigma}({\bf p}) 
+ 
\frac{{\bf p}\cdot{\bf q}}{m}
\left [ g \delta n_{-\sigma} + U_{\rm ext} ({\bf q}, \omega) \right ] 
\right \}
\nonumber
\\
=& 
-\frac{1}{\tau}
\frac{\partial f_{\sigma}^{0}}{\partial \varepsilon_{\sigma}^{0}}
\left [\nu_{\sigma} ({\bf p}) - \left ( A_{\sigma} + {\bf B}\cdot {\bf p} + Cp^{2} \right )\right ], 
\label{Equation112}
\end{align} 
where we omit ${\bf q}$ and $\omega$ in $\nu_{\sigma}$ and $\delta n_{\sigma}$
for simplicity.

We expand the fluctuation in terms of spherical harmonics as 
$
\nu_{\sigma}({\bf p}) 
\equiv 
\sum\limits_{l = 0}^{\infty}\sum\limits_{m = -l}^{l}
\nu_{\sigma, l}^{m} (p) P_{l}^{m}(\cos{\theta})e^{im\phi}. 
$
Multiplying Eq. (\ref{Equation112}) by $e^{-im'\phi}$ 
and integrating it over $\phi$, 
we obtain the reduced form of the linearized Boltzmann equation: 
\begin{align}
&
\sum\limits_{l = 0}^{\infty}
\frac{\partial f_{\sigma}^{0}}{\partial \varepsilon_{\sigma}^{0}}
\left [
\left ( 
\omega -\frac{pq}{m}\cos{\theta} 
\right )
\nu_{\sigma,l}^{m} P_{l}^{m} (\cos{\theta})
\right ]
\nonumber
\\
+ &
\frac{\partial f_{\sigma}^{0}}{\partial \varepsilon_{\sigma}^{0}}
\left (
\frac{pq}{m} \cos{\theta} 
\right )
[ g \delta n_{-\sigma} 
+ 
U_{\rm ext} ({\bf q}, \omega) ] \delta_{m, 0} 
\nonumber
\\
= &
-\frac{i}{\tau}
\frac{\partial f_{\sigma}^{0}}{\partial \varepsilon_{\sigma}^{0}}
\left [\sum\limits_{l = 0}^{\infty}\nu_{\sigma}^{m}(p) P_{l}^{m}(\cos{\theta}) - \left ( A_{\sigma} + {\bf B}\cdot {\bf p} + Cp^{2} \right ) \delta_{m,0}\right ]. 
\label{Equation116}
\end{align} 
We see that only the mode $m=0$ is coupled to the external potential, 
and thus we take the mode $m = 0$ which corresponds to the longitudinal wave.

When we take moments 
corresponding to the number of particles, the momentum, and the energy, 
the collision integral vanishes due to the conservation law. 
The coefficients $A_{\sigma}$, ${\bf B}$, and $C$ in the relaxation time approximation are determined from these conservation laws.
The resultant equations are: 
$\langle \nu_{\sigma, 0} \rangle  - A_{\sigma}W_{\sigma, 0} - C W_{\sigma, 2} = 0$, 
$\sum\limits_{\sigma} \left ( \langle p\nu_{\sigma, 1} \rangle - B W_{\sigma, 2} \right ) = 0$, 
and 
$\sum\limits_{\sigma}
\left ( \langle p^{2} \nu_{\sigma, 0} \rangle - A_{\sigma} W_{\sigma, 2} - C W_{\sigma, 4} \right ) = 0$. 
The function $W_{\sigma, n}$ is defined as 
\begin{align}
W_{\sigma, n} \equiv \int  \frac{d{\bf p}}{(2\pi\hbar)^3} 
p^{n} \frac{\partial f_{\sigma}^{0}}{\partial \varepsilon_{\sigma}^{0}}, 
\end{align} 
and we assume longitudinal sound ({\rm i.e.}, ${\bf B} \parallel {\bf k}$). 
We defined the moment $\langle p^{n} \nu_{\sigma, l}\rangle$ 
as 
\begin{align}
\langle p^{n} \nu_{\sigma, l}\rangle 
\equiv &
\int \frac{d{\bf p}}{(2\pi\hbar)^3} \frac{\partial f_{\sigma}^{0} }{ \partial \varepsilon_{\sigma}^{0}} p^{n} \nu_{\sigma, l}(p), 
\end{align}
where we used the notation $\nu_{\sigma, l}^{m = 0}(p) \equiv \nu_{\sigma, n}(p)$.

Multiplying Eq. (\ref{Equation116}) by $p^{n} P_{l}( \cos{\theta} )$, 
where $P_{l}^{m=0} (\cos{\theta}) \equiv P_{l} (\cos{\theta})$, 
and integrating it over ${\bf p}$, 
we obtain the moment equation, which we summarize as: 
\begin{align}
&
\left ( \omega + \frac{i}{\tau} \right ) 
\langle p^{n}\nu_{\sigma, l}\rangle
-\frac{l}{2l-1} \frac{q}{m} \langle p^{n+1}\nu_{\sigma, l-1}\rangle 
\\
& 
-\frac{l+1}{2l+3} \frac{q}{m} 
\langle p^{n+1}\nu_{\sigma, l+1}\rangle
+g\frac{q}{m}W_{n+1}  
\langle \nu_{-\sigma, 0}\rangle \delta_{l,1}
\nonumber 
\\
& 
-
\frac{i}{\tau}
\left [
\frac{W_{n}}{W_{0}}
+ 
\frac{1}{\Theta}
\left ( 
\frac{ W_{n} W_{2}^{2} }{W_{0}^{2} }
- 
\frac{ W_{n+2} W_{2} }{ W_{0} }
\right ) 
\right ]
\langle
\nu_{\sigma, 0} 
\rangle
\delta_{l,0} 
\nonumber 
\\
&
-
\frac{i}{\tau}
\frac{1}{\Theta}
\left ( 
\frac{ W_{n} W_{2}^{2} }
{W_{0}^{2} }
- 
\frac{ W_{n+2} W_{2} }{ W_{0} }
\right ) 
\langle
\nu_{-\sigma, 0} 
\rangle
\delta_{l,0} 
\nonumber 
\\
&
-
\frac{i}{\tau}
\frac{1}{\Theta}
\left (
W_{n+2}
- 
\frac{ W_{2} W_{n} }{W_{0} }
\right ) 
\left ( 
\langle
p^{2} \nu_{\uparrow, 0} 
\rangle
+
\langle
p^{2} \nu_{\downarrow, 0} 
\rangle
\right )
\delta_{l,0} 
\nonumber 
\\
&
-
\frac{i}{\tau}
\frac{W_{n+1}}{2 W_{2}}
\left ( 
\langle
p \nu_{\uparrow, 1} 
\rangle
+
\langle
p \nu_{\downarrow, 1} 
\rangle
\right )
\delta_{l,1} 
\nonumber 
\\
&
= 
- \frac{q}{m}W_{n+1} \delta_{l,1} U_{\rm ext} ({\bf q}, \omega), 
\label{MomentEquation}
\end{align} 
where we define 
$
\Theta \equiv 
2
\left (
W_{4}
-
W_{2}^{2} / W_{0}
\right )
$. 
We used $W_{n}\equiv W_{\sigma,n} = W_{-\sigma,n}$, 
assuming a population-balanced gas. 
In the absence of an external field ($U_{\rm ext}=0$), solutions of Eq. (\ref{MomentEquation}) give the frequency (or sound velocity) and damping of normal modes. Detailed calculations of the normal-mode solutions are presented in Ref.~\cite{Watabe2009}. 
From the solution of Eq. (\ref{MomentEquation}) including $U_{\rm ext}({\bf q}, \omega)$, 
we can calculate the density $\delta n_{\rm tot} = \langle \nu_{0}\rangle$, 
which can be written in terms of the density response function 
as $\delta n_{\rm tot}({\bf q}, \omega) = \chi({\bf q}, \omega) U_{\rm ext}({\bf q}, \omega)$. 

The density response function $\chi({\bf q}, \omega)$ is related to the dynamic structure factor $S({\bf q}, \omega)$ 
by the dissipation fluctuation theorem~\cite{Pines1994}. 
Using the detailed balance condition $S({\bf q}, \omega) = e^{\beta\hbar\omega} S({\bf q}, -\omega)$, 
and the relation ${\rm Im}\chi({\bf q}, \omega) = - \pi [S({\bf q}, \omega) - S({\bf q}, -\omega) ]$, 
the dynamic structure factor can be written as 
\begin{align}
S({\bf q}, \omega) = -\frac{1}{\pi} \frac{1}{1 - e^{-\beta \hbar\omega}} {\rm Im}\chi({\bf q}, \omega). 
\label{Schi}
\end{align} 
We now discuss collective modes that use 
the dynamic structure factor $S({\bf q}, \omega)$ calculated from the moment equation. 
The present moment method reproduces the excitation only in the phonon regime $q < k_{\rm F}$ 
(as discussed in a previous study~\cite{Watabe2009}), 
and hence our results are valid for phonon regimes.

\section{Dynamic Structure Factor from the Collisionless to Hydrodynamic Regime}\label{III}

\subsection{Strong coupling case}

\begin{figure}[htcp]
\includegraphics[width=9.5cm,height=9.5cm,keepaspectratio,clip]{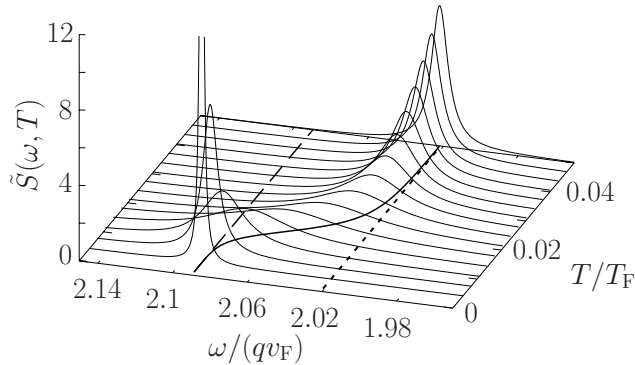}
\caption{
The dynamic structure factor $S(\omega, T)$ as a function of $\omega$ and $T$, calculated by the moment method.
In the $\omega$--$T$ plane, the solid line denotes the phase velocity calculated by the moment method~\cite{Watabe2009}, 
the dashed line gives the phase velocity of zero sound, 
and the dotted line gives that of first sound. 
}
\label{Fig1} 
\end{figure}

Figure~\ref{Fig1} shows the spectral function in a strongly coupled system as a function of $\omega$ and $T$.
We used the renormalized frequency and temperature $\omega/(v_{\rm F}q)$ and $T/T_{\rm F}$, 
where $v_{\rm F} = (\hbar/m)(3\pi^{2}N_{\rm tot}/V)^{1/3}$ is the Fermi velocity 
and $T_{\rm F} = E_{\rm F}/k_{\rm B}$ is the Fermi temperature. 
$E_{\rm F}$ is the Fermi energy, which is given by $E_{\rm F} \equiv mv_{\rm F}^{2}/2$,
$N_{\rm tot}$ is the total number of particles, which is given by $N_{\rm tot}\equiv N_{\uparrow} + N_{\downarrow}$,
and $V$ is the volume. 
The dynamic structure factor is normalized as follows: $\tilde{S}(\omega, T) \equiv {S}(\omega, T) V\varepsilon_{\rm F}/N_{\rm tot}$. 
We choose the parameters $q=0.05k_{\rm F}$ and $\alpha = 15$, 
where $\alpha \equiv g N_{\rm tot}/(V \varepsilon_{\rm F})$~\cite{comment}, 
and take moments up to $l=n=30$. 
Figure~\ref{Fig1} also shows the sound velocity calculated by the moment method (solid line), 
the zero sound (dashed line), and the first sound (dotted line) in the $\omega$--$T$ plane. 

It shows that there is a smooth crossover between zero and first sound in the structure factor.
As in Ref.~\cite{Watabe2009}, the sound eigenmode exhibits a crossover between zero and first sound. Correspondingly, the peak in the structure factor associated with the sound mode (the Brillouin peak) transitions smoothly from the zero sound mode to the first sound mode with increasing temperature.
As expected, the Brillouin peaks are quite sharp in both the collisionless and collisional regimes, 
where the damping rate is small.
The peak in the crossover regime from zero to first sound is quite broad, reflecting a short lifetime. 
The peak width in the hydrodynamic mode is discussed in Sec.~\ref{IV}. 

\begin{figure}[htpc]
\includegraphics[width=9.5cm,height=9.5cm,keepaspectratio,clip]{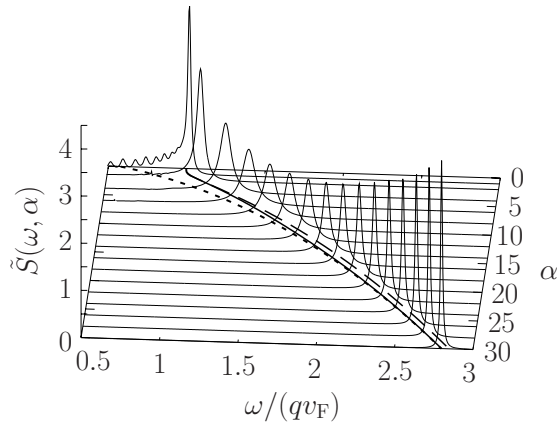}
\caption{
The dynamic structure factor $S(\omega, \alpha)$ as a function of $\omega$ and $\alpha$, calculated by the moment method. 
The sound velocity calculated by the moment method (solid line)~\cite{Watabe2009} is in the $\omega$--$\alpha$ plane. 
The sound velocities of zero sound (dashed line) and first sound (dotted line) are also plotted.
}
\label{Fig2} 
\end{figure}

Figure~\ref{Fig2} shows the spectral function 
as a function of frequency $\omega$ and coupling constant $\alpha$. 
We set the parameters as $q=0.05k_{\rm F}$ and $T= 0.025 T_{\rm F}$ and
take moments up to $l=n=31$. 
The peaks in the collisionless and collisional regimes are quite sharp. 
The Brillouin peak becomes broader in the crossover regime from zero to first sound.
The zero and first sound modes have very similar velocities in the strong-coupling regime, 
which makes it difficult to distinguish between the two regimes solely based on the mode frequency. 
However, the reduction in the intensity  of the dynamic structure factor 
with increasing coupling strength $\alpha$ 
is a clear indicator of a crossover between zero and first sound.

\subsection{Weak-coupling case}

\begin{figure}[htcp]
\includegraphics[width=9.5cm,height=9.5cm,keepaspectratio,clip]{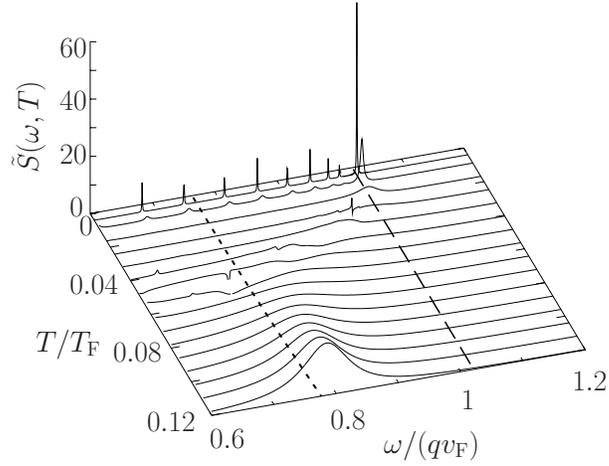}
\caption{
The dynamic structure factor $S(\omega, T)$ calculated by the moment method 
as a function of $\omega$ and $T$ for the weakly coupling case $\alpha = 1$.  
The dashed line shows the phase velocity of zero sound
and the dotted line shows that of first sound in the $\omega$--$T$ plane.
}
\label{Fig3} 
\end{figure}

The situation is more complicated for a weakly coupled system.
It is difficult to distinguish collective modes from other excitations when the moment method is used to determine the eigenvalues of a normal Fermi system
because the eigenvalue for coherent oscillation overlaps with  those for the particle--hole continuum. 
Because of this complication, 
Ref.~\cite{Watabe2009} omitted detailed discussion of the collective mode 
in a weakly coupled system with changing temperature.
We now study collective excitations in a weakly coupled system 
in terms of the dynamic structure factor. 
Figure~\ref{Fig3} shows a plot of the spectral function of a weakly coupled system. 
We choose the parameters as $q=0.01k_{\rm F}$ and $\alpha = 1$
and take moments up to $l=n=31$. 
For reference, 
the zero sound (dashed line) and first sound (dotted line) frequencies are plotted in the $\omega$--$T$ plane.

The collective modes in the collisionless and collisional regimes are easily found from the spectral intensity peak. 
The first sound mode clearly has a single broad peak at high temperature. 
Zero sound can be also identified as it has a narrow, high-intensity peak 
slightly above the Fermi velocity at low temperature.
However, in the crossover regime from zero to first sound, the intensity is low and it is affected by other modes. 

Thus, it is difficult to distinguish the collective mode in the crossover regime from zero to first sound from the structure factor in a weakly interacting system; this aspect differs from the strong coupling case.
Detailed behavior of the spectral intensity in the crossover and collisionless regimes below $\omega < q v_{\rm F}$ 
( the region corresponding to the particle--hole continuum)
is affected by the number of moments that we take for numerical calculations. 
However, the main results noted above remain the same.

\subsection{Rayleigh peak}

\begin{figure}
\includegraphics[width=9.5cm,height=9.5cm,keepaspectratio,clip]{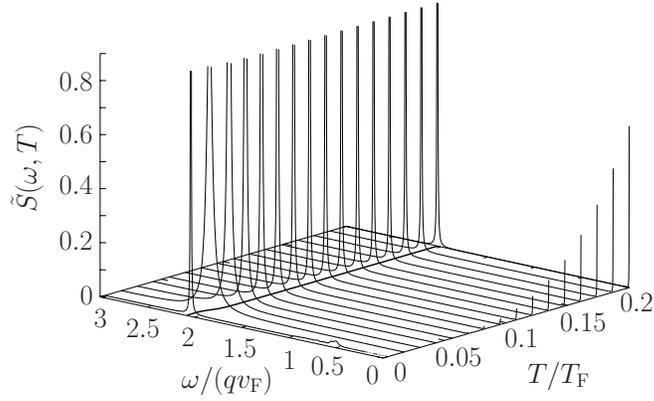}
\caption{
The dynamic structure factor $S(\omega, T)$ calculated by the moment method 
as a function of $\omega$ and $T$ using the same parameters as those for Fig.~\ref{Fig1}.
The solid line in the $\omega$--$T$ plane represents the phase velocities of the sound. 
There are two peaks: the Brillouin peak at high velocity and the Rayleigh peak near a velocity of zero.
}
\label{Fig4} 
\end{figure}

In classical hydrodynamics, 
the response function also includes a peak due to the thermal diffusion mode at $\omega=0$~\cite{Pines1994,Forster1995,Kadanoff2000}, 
which known as the Rayleigh peak~\cite{Forster1995}. 
An explicit expression for the hydrodynamic response function 
is given in Refs.~\cite{Forster1995,Kadanoff2000}; 
it is discussed in the next section 
in the context of the moment method. 
As discussed in Ref.~\cite{Watabe2009}, 
the normal mode solutions in our moment method include the thermal diffusion mode. 
The dynamic structure factor calculated for Eq. (\ref{Equation116}) involves the associated Rayleigh peak. 
Figure~\ref{Fig4} shows a plot of the spectral function over a wide range in the $\omega$--$T$ plane 
using the same parameters as those in Fig.~\ref{Fig1}. 
The Rayleigh peak appears 
at $\omega =0$ in the hydrodynamic regime, 
and it appears with increasing temperature. 
We compare this result with the hydrodynamic response function in the next section. 

\section{Dynamic Structure Factor in Hydrodynamic Regime}\label{IV}

In this section, we briefly review the density response function in the hydrodynamic regime
and we compare this result with the dynamic structure factor obtained using the moment equation. 
The average additional energy of particles in the local equilibrium $\nu_{\sigma, {\rm local}}$ is 
given by $\nu_{\sigma, {\rm local}} = A_{\sigma} + {\bf B}\cdot{\bf q} + C p^{2}$, as mentioned in Sec.~\ref{II}. 
Neglecting the departure from local equilibrium $\delta \nu_{\sigma} \equiv \nu_{\sigma} - \nu_{\sigma, {\rm local}}$ on the left-hand side of Eq.~(\ref{Equation112}) 
and also $U_{\rm ext}$, 
we obtain 
\begin{eqnarray}
\delta \nu_{\sigma} 
&=&
i \tau 
\left \{
\left ( 
\omega -\frac{{\bf p}\cdot{\bf q}}{m}
\right )
\left [
A_{\sigma} + {\bf B}  \cdot {\bf p}
+ C p^{2}
\right ]
\right . 
\nonumber
\\
&&
\left . 
\qquad
+ 
\frac{{\bf p}\cdot{\bf q}}{m}
g 
\left [ 
A_{-\sigma} 
W_{-\sigma, 0}
+
C W_{-\sigma, 2}
\right ]
\right \}. 
\label{delta-nu-ABC}
\end{eqnarray}
Substituting $\nu_{\sigma} = \nu_{\sigma, {\rm local}} + \delta \nu_{\sigma}$ in Eq.~(\ref{Equation112}) 
and taking the zeroth, first, and second moments of the linearized Boltzmann equation, 
we obtain the hydrodynamic equations given by 
\begin{align}
&
\, 
\omega
\left [ 
A_{\sigma} W_{\sigma, 0}
+ 
C W_{\sigma, 2}
\right ] 
= 
\frac{{\bf B} \cdot{\bf q}}{3m}W_{\sigma,2}, 
\\
&
\sum\limits_{\sigma}
\left \{ 
\omega{\bf B} W_{\sigma,2}
-
A_{\sigma} \frac{W_{\sigma,2}}{m}{\bf q}
-
C \frac{W_{\sigma,4}}{m}{\bf q}
\right . 
\nonumber 
\\
& 
\qquad \qquad \qquad 
\left .
+ 
g \frac{W_{\sigma,2}}{m}
\left [
A_{-\sigma}  W_{-\sigma,0} 
+ 
C W_{-\sigma,2}
\right ]
{\bf q}
\right \} 
\nonumber
\\
&
\qquad \qquad  \qquad
-
i4\eta q^{2} 
{\bf B} 
+ \frac{\bf q}{m} (W_{\uparrow,2} + W_{\downarrow,2}) U_{\rm ext} 
= 0, 
\\
&
\omega 
\left [
A_{\uparrow}  W_{\uparrow,2}
+ 
A_{\downarrow}  W_{\downarrow,2}
+ 
C 
(W_{\uparrow,4}+W_{\downarrow,4})
\right ] 
\nonumber
\\
&
\qquad \qquad \qquad
-
\frac{{\bf B} \cdot{\bf q}}{3m}
(W_{\uparrow,4}+W_{\downarrow,4})
- 
4i\kappa m^{2} C T q^{2}
=0, 
\end{align} 
where the collision integral is zero in the hydrodynamic regime because of the conservation law. 
$\kappa$ is the thermal conductivity $\kappa = - k_{\rm B}\beta \tau (W_{6} - W_{4}^{2}/W_{2})/(6m^{4})$, 
and $\eta$ is the shear viscosity $\eta = - 2\tau W_{4}/(15m^{2})$. 
Detailed derivations are given in Ref.~\cite{Watabe2009}.

The above equations can be written in terms of fluctuations in 
the total density $\delta n_{\rm tot} \equiv \sum_{\sigma} \left [ A_{\sigma} W_{\sigma,0} + C W_{\sigma, 2} \right ]$, 
velocity $\overline{\bf v} \equiv - {\bf B}$, and 
energy $\delta E = \sum_{\sigma} \left [ A_{\sigma} W_{\sigma, 2} +  C W_{\sigma, 2} \right]/(2m)$. 
The resultant hydrodynamic equations can be expressed in matrix form: 
\begin{align}
K
\begin{pmatrix}
\delta n_{\rm tot}  \\
{\bf q}\cdot \overline{\bf v} \\
\delta E 
\end{pmatrix} 
= 
\begin{pmatrix}
0 \\
 \displaystyle{\frac{q^{2}}{m}  U_{\rm ext}  }\\
0
\end{pmatrix}, 
\label{Matrix14}
\end{align}
where the 3$\times$3 matrix $K$ is given by 
\begin{align} 
K \equiv 
\begin{pmatrix}
\omega & \displaystyle{\frac{2W_{2}}{3m}} & 0 \\
-\displaystyle{\frac{gq^{2}}{2m}} & \displaystyle{\omega - i\frac{2\eta q^{2}}{W_{2}} }& \displaystyle{\frac{q^{2}}{W_{2}}} \\
\displaystyle{-i\frac{\Gamma_{\kappa} \gamma W_{2}}{2m W_{0}} }& \displaystyle{\frac{W_{4}}{3m^{2}} }& \omega + i \Gamma_{\kappa} \gamma
\end{pmatrix}. 
\label{Matrix15}
\end{align}
The parameter $\gamma$ is defined by 
\begin{align}
\gamma \equiv \frac{W_{0}(W_{4}-gW_{2}^{2})}{W_{2}^{2}(1-gW_{0})}, 
\end{align}
and the rate $\Gamma_{\kappa}$ is given by 
\begin{align}
\Gamma_{\kappa} \equiv - \frac{2 \kappa T m^{2} q^{2} W_{2}^{2} (1-gW_{0})}{(W_{4}-gW_{2}^{2})(W_{4}W_{0}-W_{2}^{2})}. 
\label{Gammak}
\end{align} 
Solving the matrix equation (\ref{Matrix14}) for $\delta n_{\rm tot}$, 
we obtain 
\begin{align}
\delta n_{\rm tot} 
& = \frac{1}{{\rm det}K} (K_{13}K_{32} - K_{12}K_{33}) \frac{q^{2}}{m} U_{\rm ext}
\\
& = - \frac{2W_{2}}{3m^{2}}q^{2} \frac{\omega + i\Gamma_{\kappa} \gamma}{{\rm det}K} U_{\rm ext}. 
\end{align} 
We neglect the second-order terms in the transport coefficients $\kappa$ and $\eta$ since they are small in the hydrodynamic regime.
This allows the determinant of the matrix $K$ to be reduced to 
${\rm det}K = (\omega^{2} - \Omega^{2}) ( \omega + i \Gamma_{\kappa} ) + 2 i \Gamma \omega^{2}$, 
where 
\begin{align}
\Omega  
\equiv  
\sqrt{\frac{W_{4}-g W_{2}^{2}}{3W_{2}}}\frac{q}{m} \equiv c q
\label{ReFirstSound}
\end{align}
and 
\begin{align}
\Gamma = - \frac{\eta q^{2}}{W_{2}} - \frac{\kappa T q^{2} m^{2}}{(W_{4}-gW_{2}^{2})}. 
\end{align}
Taking this determinant to be zero, 
we obtain the eigenmodes of the hydrodynamic modes (to first order in $\kappa$ and $\eta$): 
$\omega = \pm \Omega - i \Gamma$ and $\omega = -i\Gamma_{\kappa}$~\cite{Watabe2009}. 
$\Omega$ is the frequency of the sound mode and $\Gamma$ is its damping rate. 
$c$ in Eq.~(\ref{ReFirstSound}) is the sound velocity. 
The damping rate $\Gamma_{\kappa}$ is that of the heat diffusion mode.

Using the above approximations for the zeros of ${\rm det}K$, 
we can explicitly derive the density disturbance induced by the external field 
in terms of the frequency and damping rates of the eigenmodes. 
This is given as 
$\delta n_{\rm tot} ({\bf q}, \omega)  = \chi ({\bf q}, \omega) U_{\rm ext} ({\bf q}, \omega) $, 
where the resultant density response function $\chi ({\bf q}, \omega) $ is given by 
\begin{align}
\chi ({\bf q}, \omega) = - \frac{2W_{2}}{3m^{2}}q^{2} \frac{\omega+i\Gamma_{\kappa} \gamma}
{(\omega - \Omega + i\Gamma)(\omega + \Omega + i\Gamma)(\omega + i\Gamma_{\kappa})}.  
\end{align}
The absorptive susceptibility can be reduce to 
\begin{align}
{\rm Im}\chi ({\bf q}, \omega) 
= & \frac{2W_{2}}{3m^{2}c^{2}} 
\left [ 
\frac{\omega (\gamma-1)\Gamma_{\kappa}}{\omega^{2}+\Gamma_{\kappa}^{2}}
+ 
\frac{2\omega\Gamma\Omega^{2}}{(\omega^{2}-\Omega^{2})^{2} + (2\omega\Gamma)^{2}}
\right . 
\nonumber
\\
&
\left .
-
\frac{\omega\Gamma_{\kappa} (\gamma-1)(\omega^{2}-\Omega^{2})}{(\omega^{2}-\Omega^{2})^{2} + (2\omega\Gamma)^{2}}
\right ], 
\label{SQWHDE}
\end{align} 
where we used $\Omega \gg \Gamma_{\kappa}$, and also $\Omega \gg \Gamma$. 
This absorptive susceptibility has two peaks: 
the Rayleigh peak at $\omega = 0$ and the Brillouin peak at $\omega = \Omega$. 

\begin{figure}[tcp]
\includegraphics[width=9.5cm,height=9.5cm,keepaspectratio,clip]{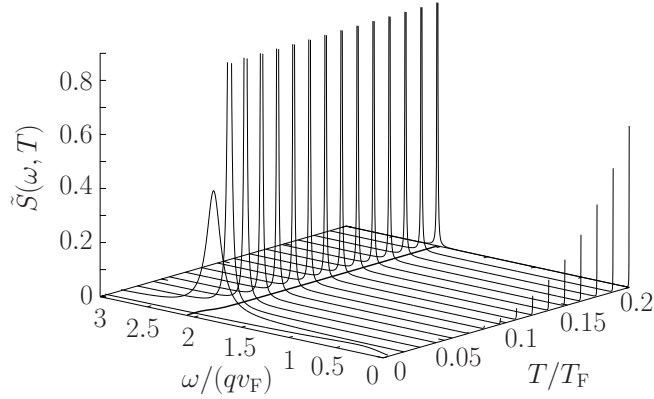}
\caption{
The dynamic structure factor $S(\omega, T)$ obtained from the hydrodynamic equations given in Eq.~(\ref{SQWHDE}), 
using the same parameters as those for Fig.~\ref{Fig1}. 
}
\label{Fig5} 
\end{figure}

Figure~\ref{Fig5} plots the dynamic structure factor obtained from the hydrodynamic equations as a function of $\omega$ and $T$.
We used the same parameter set as that in Fig.~\ref{Fig1}. 
Comparing Fig.~\ref{Fig5} with Fig.~\ref{Fig4} shows that 
the structure factor calculated using the hydrodynamic equations 
differs from that obtained by the moment method in the crossover and collisionless regimes, as expected.
The spectral intensity decreases in those regimes because of the high damping rates of hydrodynamic modes. 
In contrast, Fig.~\ref{Fig4} clearly shows 
that the moment method can correctly describe the dynamic structure factor
in both the crossover and collisionless regimes, whereas the hydrodynamic equations cannot.

\begin{figure}[tcp]
\includegraphics[width=7cm,height=7cm,keepaspectratio,clip]{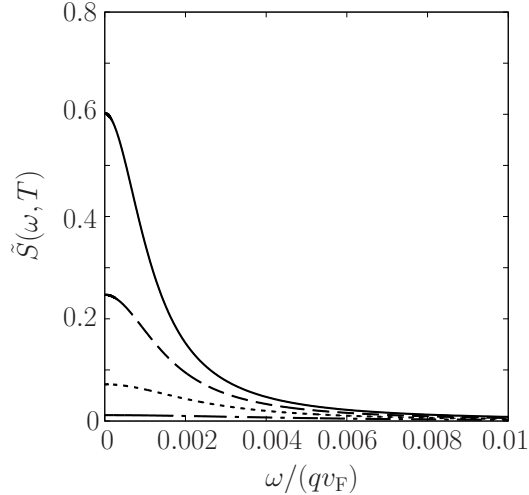}
\caption{
Rayleigh peaks for several temperatures as a function of $\omega$. 
Solid, dashed, dotted, and dot-dashed lines represent 
results for $T/T_{\rm F} = 0.2$, $0.1608$, $0.1216$, and $8.24 \times 10^{-2}$, respectively. 
}
\label{Fig6} 
\end{figure}

To evaluate how effective the moment method is in the collisional regime,
we perform a more quantitative comparison between 
the results obtained using the moment method and those obtained using the hydrodynamic equations.
Figure~\ref{Fig6} shows a plot of Rayleigh peaks for several temperatures as a function of $\omega$.
The solid, dashed, dotted and dot-dashed lines represent 
results for $T/T_{\rm F} = 0.2$, $0.1608$, $0.1216$, and $8.24 \times 10^{-2}$, respectively. 
The thick lines are the results obtained using the moment method and the thin lines are those obtained using the hydrodynamic equations.
It is not possible to distinguish them 
since they exhibit identical behavior in the hydrodynamic regime. 
At lower temperatures, the Rayleigh peak becomes too weak to distinguish the two results.

\begin{figure}[tcp]
\includegraphics[width=7cm,height=7cm,keepaspectratio,clip]{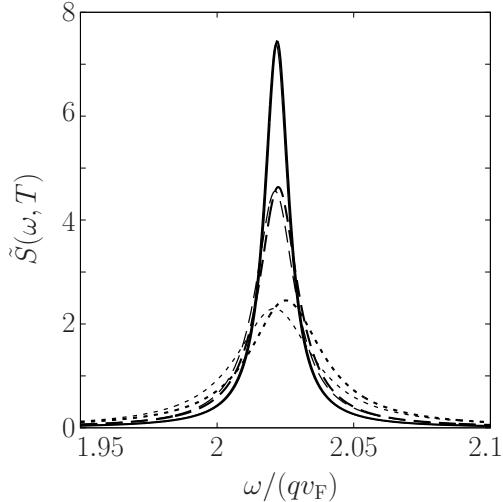}
\caption{
Brillouin peaks for several temperatures as a function of $\omega$.  
Solid, dashed, and dotted lines represent 
results for $T/T_{\rm F} = 5 \times 10^{-2}$, $4.02 \times 10^{-2}$, and $3.04 \times 10^{-2}$, respectively. 
Thick lines are the results obtained by the moment method and the thin lines are the results obtained from the hydrodynamic equations.
}
\label{Fig7} 
\end{figure}

Figure~\ref{Fig7} shows a plot of Brillouin peaks for several temperatures as a function of $\omega$.
The solid, dashed, and dotted lines represent the 
results for $T/T_{\rm F} = 5 \times 10^{-2}$, $4.02 \times 10^{-2}$, and $3.04 \times 10^{-2}$, respectively.
The thick lines show the results obtained using the moment method and the thin lines show the results obtained using the hydrodynamic equations.
As the temperature decreases and the system approaches the crossover regime, 
the difference between the dynamic structure factor obtained by the moment method and 
that obtained from the hydrodynamic equations increases.
In both Figs.~\ref{Fig6} and~\ref{Fig7}, 
we used the same parameters as in Fig.~\ref{Fig1}. 
We conclude that our moment method can reproduce the dynamic structure factor in the hydrodynamic regime reasonably well.

\begin{figure}[tcp]
\includegraphics[width=7cm,height=7cm,keepaspectratio,clip]{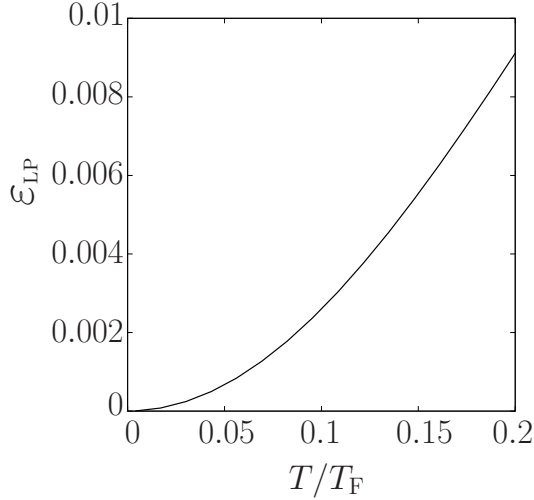}
\caption{
The Landau--Placzek ratio $\varepsilon_{\rm LP}$ obtained from our formalism as a function of temperature. 
}
\label{Fig8} 
\end{figure}

We comment on the Landau--Placzek ratio in our formulation. 
Comparing Eq. (\ref{SQWHDE}) 
with the formula for the absorptive susceptibility in terms of the thermodynamic quantities 
in Eq. (4.44a) in Ref.~\cite{Forster1995} 
(also Eq. (87a) in Ref.~\cite{Kadanoff2000}), 
we find that $C_{p}/C_{v}$ is equal to $\gamma$, 
where $C_{p}$ and $C_{v}$ are the specific heats at constant pressure and volume, respectively. 
We used $n_{\rm tot} = 2n_{\sigma} = -2W_{2}/(3m)$ and the relation 
$mc^{2} = (\partial p / \partial n_{\rm tot})|_{S} = (C_{p}/C_{v})(\partial p / \partial n_{\rm tot})|_{T}$, 
where $p$ is the pressure and $S$ is the entropy. 
The Landau--Placzek ratio $\varepsilon_{\rm LP}$, 
defined as the ratio of half the area under the Rayleigh peak to that under one Brillouin peak~\cite{Forster1995}, 
is given by $\varepsilon_{\rm LP} \equiv (C_{p}/C_{v})-1$~\cite{Landau1934}; 
hence the Landau--Placzek ratio can be obtained as 
\begin{align}
\varepsilon_{\rm LP} & = \gamma-1 
\\
&
= \frac{W_{0}W_{4}-W_{2}^{2}}{W_{2}^{2}(1-gW_{0})}. 
\end{align}
At $T = 0$, we have $\chi_{n,\sigma} = - 3N_{\rm tot}p_{\rm F}^{n}/(4V\varepsilon_{\rm F})$, 
and hence the Landau--Placzek ratio $\varepsilon_{\rm LP}$ becomes zero. 
Figure~\ref{Fig8} shows that the temperature dependence of the Landau--Placzek ratio for $\alpha = 15$. 
As mentioned above, this ratio approaches zero at $T = 0$
and it is a monotonically increasing function of temperature. 
To derive the Landau--Placzek ratio, it is usually necessary
to calculate the specific heats at constant pressure and volume $C_{p}$ and $C_{v}$. 
However, in our formalism it can simply be obtained from the function $W_{n}$.

\section{Discussion}\label{V}

In experiments of ultracold atomic gases, 
two-photon Bragg scattering 
is a well-developed tool for studying the dynamic structure factor~\cite{Veeravalli2008}. 
This type of study is analogous to the classic work by Abrikosov and Khalatnikov~\cite{Abrikosov1958}, who 
proposed using light scattering to observe zero sound in liquid $^{3}$He. 
With the recent dramatic progress in experimental techniques for ultracold atomic gases, 
the coupling strength $\alpha$ can also be controlled because of the Feshbach resonance. 
The results in the present paper should be observable in principle. 

\begin{figure}[tcp]
\includegraphics[width=9.5cm,height=9.5cm,keepaspectratio,clip]{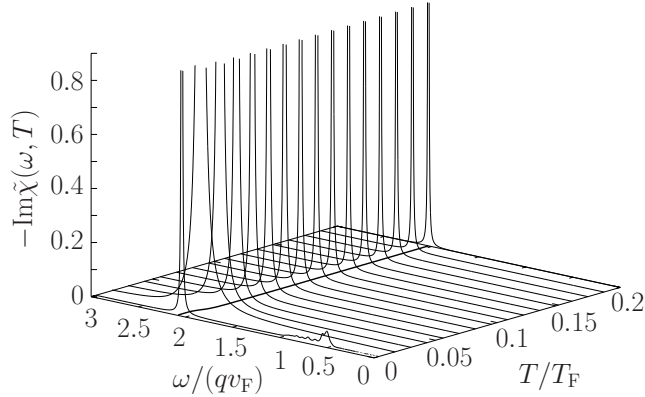}
\caption{
The imaginary part of the density response function $-{\rm Im}\chi (\omega, T)$ 
as a function of $\omega$ and $T$ using the same parameters as for Fig.~\ref{Fig1}. 
The solid line in the $\omega$--$T$ plane shows the phase velocity of the sound 
obtained by the moment method. 
Only the Brillouin peak is visible at high velocity.
The Rayleigh peak near zero velocity is not visible. 
}
\label{Fig9} 
\end{figure}

Hu {\it et al}. recently discussed 
the density response function in superfluid gases in the two-fluid hydrodynamic regime~\cite{Hu2010}. 
They showed that 
the first and second sound have different relative weights in the dynamic structure factor $S({\bf q}, \omega)$ 
and the density response function ${\rm Im}\chi ({\bf q}, \omega)$~\cite{Hu2010}. 
This is clearly seen in the relation between both functions shown in Eq. (\ref{Schi}). 
When $\omega/(k_{\rm B}T) \ll 1$, 
$S({\bf q}, \omega) \simeq - k_{\rm B}T{\rm Im}\chi ({\bf q}, \omega)/(\pi\hbar\omega)$. 
Hu {\it et al}. pointed out that the extra factor $1/\omega$ leads to a large enhancement 
of the peak associated with the low-frequency second sound in $S({\bf q}, \omega)$~\cite{Hu2010}. 
This is also true for the Rayleigh peak in the present study. 
Figure~\ref{Fig9} plots the imaginary part of the density response function $-{\rm Im}\chi (\omega, T)$ 
as a function of $\omega$ and $T$ using the same parameters as for Fig.~\ref{Fig1}.
The density response function is normalized by 
$\tilde{\chi}(\omega, T) \equiv {\chi}(\omega, T) V\varepsilon_{\rm F}/N_{\rm tot}$. 
Due to the absence of the factor $1/\omega$, 
the Rayleigh peak near $\omega = 0$ cannot be observed. 
Thus, the Rayleigh peak can be observed through the dynamic structure factor, 
whereas it has a negligibly small weight in the density response function. 
A localized potential that turns off after a short duration, 
which is used in experimental studies of ultracold atomic gases, 
will also be useful for studying the Rayleigh peak 
because the excitation of density pulses is proportional to ${\rm Im}\chi({\bf q}, \omega)/\omega$ \cite{Arahata2009,Hu2010}. 

The moment method can be applied to other systems, 
such as transverse zero sound and a collective mode in a Fermi gas with dipole interaction. 
The spectral function of a polarized normal Fermi gas at the unitarity limit 
has recently been studied at $T=0$~\cite{Stringari2009}. 
The method described in the present paper can also be used to determine the density response of a polarized normal Fermi gas with a strong interaction at finite temperatures.
We intend to apply the moment method to these systems in the future.

In the present paper, we applied the moment method to the Boltzmann equation with a simple mean-field interaction and calculated the dynamic structure factor in the crossover regime between the collisionless regime and the hydrodynamic regime. 
Our method provides a controlled calculational tool as long as the perturbative control of the interaction is possible. 
In the case of a normal Fermi gas near the unitarity limit, 
the Boltzmann equation (\ref{Boltzmann}) will be modified 
by replacing the mean-field interaction $U_{\sigma} = g n_{-\sigma}$ (with $g = 4\pi\hbar^{2}a/m$) 
with the real part of self-energy~\cite{KadanoffBaym} associated with the many-body $T$ matrix ${\mathcal T}$ 
which is calculated through the ladder approximation~\cite{Bruun2004, Chiacchiera2009}. 
For the collision integral in the Boltzmann equation, 
the scattering cross section $d\sigma/d\Omega$ is also given in terms of the many-body $T$ matrix ${\mathcal T}$ 
through $d\sigma/d\Omega = m^{2} |{\mathcal T}|^{2} /(4\pi\hbar^{2})^{2}$~\cite{Bruun2005, Riedl2008,Chiacchiera2009}. 

The above approach involves difficulty when one solves this Baltzmann equation directly; 
but, one can still apply the moment method if one uses the so-called unitarized vacuum scattering matrix, 
{\it i.e.} the $T$ matrix neglecting the effects of the medium given by ${\mathcal T} = g/(1+ia p/\hbar)$, 
where $p$ is 
the relative momentum of the scattering particles~\cite{Riedl2008, Bruun2005, Massignan2005, Chiacchiera2009, Bruun2007}. 
This approach using the vacuum scattering 
is valid for the high temperature regime $(T - T_{\rm c})/T_{\rm c} \gtrsim 1$~\cite{Riedl2008}. 
The effects of the medium becomes significant at lower temperatures $(T - T_{\rm c})/T_{\rm c} \lesssim 1$~\cite{Bruun2005,Riedl2008} 
because of the phase transition associated with the Cooper instability~\cite{Bruun2005}. 
We note that the regime $T  \lesssim 2 T_{\rm c}$ near the unitarity limit corresponds to the pseudogap regime~\cite{Bruun2005, Riedl2008, Watanabe2010}. 
As a result, the moment method will be applicable for the normal Fermi gas near the unitarity limit at higher temperature than that for the pseudogap regime. 
Outside the perturbative regime, the results may be trusted qualitatively but only to the same extent that one can trust qualitative descriptions of the mean-field theory for static equilibrium properties.


\section{Conclusion}\label{VI}
We studied the spectral function of a normal Fermi system 
at finite temperatures from the collisionless to the hydrodynamic regime. 
We solved the Boltzmann equation accurately using the moment method
and we determined the dynamic structure factor in the crossover and collisionless regimes 
as well as in the hydrodynamic regime 
as a function of the temperature and coupling strength. 
We compared the results obtained by the moment method with those obtained using the hydrodynamic equations
in terms of the Rayleigh and Brillouin peaks. 
We also briefly commented on the Landau--Placzek ratio. 

It has been generally difficult to study the dynamic structure factor 
in the crossover regime between the collisionless and hydrodynamic regimes because of its complexity. 
This study describes a powerful method for calculating the dynamic structure factor 
over the whole crossover range, from the collisionless to the hydrodynamic regime, within a single framework.

\section{acknowledgment}
S. W. acknowledges support from a Grant-in-Aid for JSPS Fellows (217751). 
T. N. was supported by Grant-in-Aid for Scientific Research from JSPS.


\begin{thebibliography}{99}
\bibitem{comment}
The coupling strength $\alpha$ is related to the $s$-wave scattering length $a$ through $\alpha \equiv 8k_{\rm F}a/(3\pi)$ 
where $k_{\rm F}$ is the Fermi wavenumber given by $k_{\rm F} \equiv (3\pi^{2}N_{\rm tot}/V)^{1/3}$.
\end{thebibliography}
\end{document}